\documentclass[sigconf, nonacm]{acmart}

\usepackage{tikz}
\usepackage{amsmath}

\usepackage{multirow}
\usepackage{multicol}
\usepackage{booktabs}
\usepackage{makecell}
\usepackage[binary-units=true,group-separator={,},group-minimum-digits={3}]{siunitx}
\usepackage{paralist}
\usepackage{numprint}
\usepackage{comment}
\usepackage{array}
\usepackage{balance}
\usepackage[inline]{enumitem}
\usepackage{color, colortbl}
\usepackage{xcolor}
\usepackage{numprint}
\usepackage{array}
\usepackage{graphicx}
\usepackage{lipsum}
\usepackage{pgfplots}
\usepackage{subcaption}
\usepackage[linesnumbered,lined,boxed,commentsnumbered,noend]{algorithm2e}
\usepackage{nicefrac}
\usepackage{url}
\usepackage{fancyvrb}
\usepackage{listings}
\usepackage{microtype}
\usepackage{tikz}
\usepackage{appendix}
\usepackage{adjustbox}

\usepackage{caption}
\usepackage{comment}

\usepackage{pifont}

\SetCommentSty{mycommfont}

\pgfplotsset{compat=1.15}
\usepgfplotslibrary{colorbrewer}
\usepgfplotslibrary{colormaps}
\pgfplotsset{grid style={dotted,gray}}
\usetikzlibrary{tikzmark}

\usepackage{float} 

\newcommand\revision[1]{\textcolor{red}{#1}}
\newcommand{\papertitle}{\textsc{Blink}}
\newcommand{\hibench}{\textsc{HiBench}}
\newcommand{\hadoop}{\textsc{Hadoop MapReduce}}

\newcommand{\mapreduce}{\textsc{MapReduce}}
\newcommand{\apache}{\textsc{Apache}}
\newcommand{\spark}{\textsc{Spark}}
\newcommand{\flink}{\textsc{Flink}}
\newcommand{\storm}{\textsc{Storm}}
\newcommand{\sparkfull}{\textsc{Apa\-che Spark}}
\newcommand{\flinkfull}{\textsc{Apa\-che Flink}}
\newcommand{\stormfull}{\textsc{Apa\-che Storm}}
\newcommand{\mllib}{\textsc{MLlib}}
\newcommand{\rdd}{\textsc{rdd}}

\newcommand{\doe}{\textsc{DoE}}
\newcommand{\DAG}{\textsc{dag}}

\newcommand{\ernest}{\textsc{Ernest}}
\newcommand{\bnumber}{\textsc{block-n}}
\newcommand{\bsize}{\textsc{block-s}}
\newcommand{\dataset}{\textsc{d}}

\newcommand{\cmark}{\ding{51}}%
\newcommand{\xmark}{\ding{55}}%

\newcommand\vldbdoi{XX.XX/XXX.XX}
\newcommand\vldbpages{XXX-XXX}
\newcommand\vldbvolume{14}
\newcommand\vldbissue{1}
\newcommand\vldbyear{2020}
\newcommand\vldbauthors{\authors}
\newcommand\vldbtitle{\shorttitle} 
\newcommand\vldbavailabilityurl{URL_TO_YOUR_ARTIFACTS}
\newcommand\vldbpagestyle{plain} 

\begin{document}
\title{Blink: Lightweight Sample Runs for Cost Optimization of Big Data Applications}


\author{Hani Al-Sayeh}
\affiliation{%
  \institution{TU Ilmenau, Germany}
}
\email{hani-bassam.al-sayeh@tu-ilmenau.de}

\author{Muhammad Attahir Jibril}
\affiliation{%
  \institution{TU Ilmenau, Germany}
}
\email{muhammad-attahir.jibril@tu-ilmenau.de}

\author{Bunjamin Memishi}

\affiliation{%
  \institution{Riinvest College, Kosovo}
}
\email{bunjamin.memishi@riinvest.net}

\author{Kai-Uwe Sattler}
\affiliation{%
  \institution{TU Ilmenau, Germany}
}
\email{kus@tu-ilmenau.de}

\begin{abstract}

Distributed in-memory data processing engines accelerate iterative applications by caching substantial datasets in memory rather than recomputing them in each iteration.
Selecting a suitable cluster size for caching these datasets plays an essential role in achieving optimal performance.
In practice, this is a tedious and hard task for end users, who are typically not aware of cluster specifications, workload semantics and sizes of intermediate data.

We present \papertitle{}, an autonomous sampling-based framework, which predicts sizes of cached datasets and selects optimal cluster size without relying on historical runs.
We evaluate \papertitle{} on a variety of iterative, real-world, machine learning applications.
With an average sample runs cost of \SI{4.6}{\percent} compared to the cost of optimal runs, \papertitle{} selects the optimal cluster size in 15 out of 16 cases, saving up to \SI{47.4}{\percent} of execution cost compared to average costs.


\end{abstract}

\maketitle

\section{Introduction}
\label{sec:introduction}

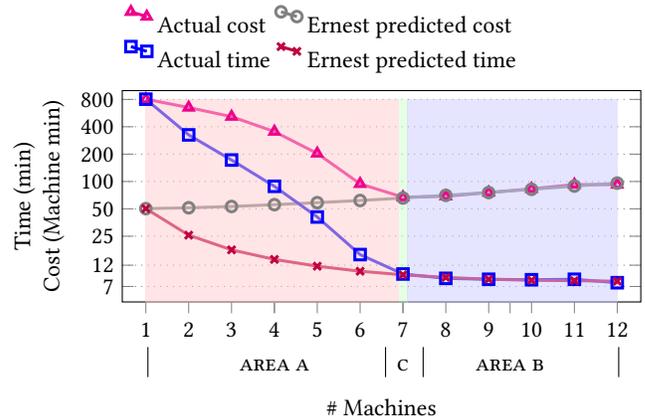
\begin{figure}[t!]
\ref*{svm-intro}
\centering
\begin{tikzpicture}

\begin{semilogyaxis}[ybar,
             bar width=4pt,
             xtick=data,
             ymin=0,
             xlabel=\# Machines,
             ylabel style={align=center},
             ylabel=Time (min)\\Cost (Machine min),
             width=1.0\columnwidth,
             enlarge x limits = 0.05,
             ymin = 0,
             ymax = 1000,
             ytick = {1, 7, 12, 25, 50, 100, 200, 400,  800},
             yticklabels = {1, 7, 12, 25, 50, 100, 200, 400, 800},
             xtick pos=bottom,
             height=125pt,
             table/col sep=comma,
             table/x=Machines,
             legend cell align=left,
             ymajorgrids,
             cycle list/Blues-3,
             every axis plot/.append style={},
             legend style={draw=none},
             legend columns=2,
             legend to name=svm-intro,
             extra x ticks={4,7,9.5},
             extra x tick labels={\textsc{area a}, \textsc{c}, \textsc{area b}},
             extra x tick style={
             tick label style={yshift=-5mm}
             }
             ]
\addplot[line width=0.4mm, magenta,sharp plot, mark=triangle] table [y=ActualCost] {data/svm-intro.csv};
\label{plot:actual_cost:actual_cost}
\addplot[line width=0.4mm, gray,sharp plot, mark=o] table [y=ErnestPredictedCost] {data/svm-intro.csv};
\label{plot:ernest_predicted_cost:ernest_predicted_cost}
\addplot[line width=0.4mm, blue,sharp plot, mark=square] table[y=ActualLatency] {data/svm-intro.csv}; 
\label{plot:actual_latency:actual_latency}
\addplot[line width=0.4mm, purple,sharp plot, mark=x] table[y=ErnestPredictedLatency] {data/svm-intro.csv}; 
\label{plot:ernest_latency:ernest_latency}

\draw [draw=none, fill=red!20, semitransparent] (1,0) rectangle (6.9,800);
\draw [draw=none, fill=green!20, semitransparent] (6.9,0) rectangle (7.1,800);
\draw [draw=none, fill=blue!20, semitransparent] (7.1,0) rectangle (12,800);

\legend{Actual cost, Ernest predicted cost, Actual time, Ernest predicted time}
\end{semilogyaxis}

\draw (0.341,-0.6) -- (0.341,-1);
\draw (3.5,-0.6) -- (3.5,-1);
\draw (4.0,-0.6) -- (4.0,-1);
\draw (6.6,-0.6) -- (6.6,-1);

\end{tikzpicture}
\vspace{-20pt}
\caption{Selection of suitable cluster configuration (\textsc{svm})}
\label{plot:svm-intro}
\vspace{-15pt}
\end{figure}

With the recent advent of compute-intensive, iterative machine learning applications, modern distributed systems such as Spark~\cite{Zaharia:2010:SCC:1863103.1863113}, Storm~\cite{Apache:Storm}, and Flink~\cite{Carbone:2015:DEB:ApacheFlink} enhance the performance of such applications by caching crucial datasets in memory instead of recomputing or fetching them from slower storage (e.g., disk or HDFS) in each iteration~\cite{Yu:2017:INFOCOM:LRC}. 
To measure the impact of repetitive re-computations on system performance, we run \emph{Support Vector Machine} (\textsc{svm}) application on an input dataset of \SI{59.5}{\giga\byte} using different cluster sizes (\numrange{1}{12} machines) on our private cluster (cf.~Section~\ref{sec:evaluation}). 
We measure the actual execution time and the cost (\#machines $\times$ time) of each application run. As depicted in Figure~\ref{plot:svm-intro}, we distinguish three areas:

\begin{itemize}
    \item Area A (\numrange{1}{7} machines): Increasing the cluster size decreases both execution time and cost.
    \item Area B (\numrange{7}{12} machines): Increasing the cluster size decreases execution time but increases execution cost.
    \item Area C (7 machines): The junction of both areas, where the highest cost efficiency is achieved.
\end{itemize}

\noindent
In area A, the total memory capacity of the cluster machines is not enough for caching all partitions of a certain crucial dataset in \textsc{svm}. 
As a result, many of its partitions do not fit in memory and are re-computed in all iterations, which is very expensive.
A deeper dive into a single iteration shows that:
\begin{enumerate*}
    \item The percentage of cached data partitions in area A for 7 to 1 machines are \SI{100}{\percent}, \SI{92}{\percent}, \SI{87}{\percent}, \SI{70}{\percent}, \SI{52}{\percent}, \SI{35}{\percent} and \SI{17}{\percent}, respectively.
    \item On average, a task that reads an already cached partition runs 97$\times$ shorter than a task that recomputes a 
    partition of equal size.
\end{enumerate*}
On the other hand, in area B, increasing the cluster size reduces the execution time of the parallel part of the application but does not influence the serial part~\cite{Amdahl:1967:AmdahlLaw}. Also, there is an additional overhead of transferring data between more machines.
All these decrease cost efficiency.

Even though area B reduces the system latency compared to area C, area C is considered as the optimal cluster size for the following reasons.
First, in public clouds where a pay-as-you-go pricing model\cite{gohad:2013:CCEM:cloud, narayanan:2020:DISPA:analysis} is used, running applications in area C minimizes monetary costs.
Second, in resource-constrained private clouds, running applications in area C is a better utilization of these limited resources, resulting in increased system throughput. 
In area A, resources are wasted in repetitive re-computations and in area B, they are wasted while processing the serial part of the application and transferring datasets between many machines.
Third, adding more machines reduces the application execution time (to some extent) in a non-linear way~\cite{Amdahl:1967:AmdahlLaw} while the cost increases linearly (see Figure~\ref{plot:svm-intro}). 
In other words, the optimum cost in area C is not at the expense of latency, as there is a correlation between the two ($ Cost = \#Machine \times Time $).
Fourth, minimizing the execution cost does not necessarily require runtime prediction.
To run an application in area C, we just need to know the size of cached datasets and caching capacity of each machine. 
This is more adaptive to cluster changes (different machine/instance types) and much simpler than other runtime prediction approaches that require many experiments and are influenced by several configurations and system dynamics (concurrent jobs interference~\cite{wang:2016:ICCC:modeling}, application parameters~\cite{Alsayeh:2020:DAPD:Gray}, +200 Spark configurations~\cite{Apache:SparkConf}, stragglers~\cite{ousterhout:2015:NSDI:making}, operating system and JVM uncertainties~\cite{Alipourfard:2017:NSDI:CherryPick}, etc.).

Runtime prediction approaches carry out sample runs on small datasets to predict the execution time of the \textit{actual run}, which processes the original huge input data. 
The challenging part of these approaches (besides the non-tolerable overhead of sample runs) comes when the sample datasets fit into memory and the original input data does not, which is most probably the case.
In this case, these approaches predict the execution time accurately only if cache evictions do not take place in actual runs.
We make predictions for the \textsc{svm} experiments (cf. Figure~\ref{plot:svm-intro}) using Ernest~\cite{Venkataraman:2016:NSDI:Ernest} (cf. Section~\ref{sec:related-work} for details) and realize that its prediction is accurate only in area B.
But since its runtime model does not factor in memory limitation, its prediction is inaccurate in area A.
Even worse, Ernest predicts that a single machine cluster size leads to minimal cost. 
However, the actual cost on a single machine is higher than the optimal cost (on 7 machines) by 12$\times$ and Ernest's prediction by 16$\times$ (cf. Figure~\ref{plot:svm-intro}).

End users (or sometimes online schedulers) need additional support in selecting cloud configurations to run their (black-box) application binaries (e.g., jar files). 
Typically, they are not aware of application semantics, sizes of cached datasets, and cluster specification and it becomes crucial when historical runs and statistics are missing (more than 60\% of jobs running in data production clusters are non-recurring~\cite{agarwal2012reoptimizing,Alipourfard:2017:NSDI:CherryPick,ferguson2012jockey}).

To tackle this problem, we need to predict the total size of cached datasets and, based on it, we deduce the amount of memory required for an eviction-free execution and, in turn, the cluster size.
Since historical statistics are not available for non-recurring applications~\cite{Venkataraman:2016:NSDI:Ernest, agarwal2012reoptimizing, jyothi2016morpheus, al:2020:DISPA:masha}, low overhead sample runs on a small amount of data are required.
To this end, we present \papertitle{}.

\papertitle{} is an autonomous sampling-based framework that performs optimal resource provisioning for big data iterative applications.
\papertitle{} performs lightweight sample runs with sample data size in the range of \SIrange{0.1}{0.3}{\percent} of the complete input data scale -- and even lower.
By analyzing these sample runs, \papertitle{} predicts the size of cached datasets and the memory footprint of the application during the actual run.
Then, it selects an optimal cluster size (area C) that provides the required amount of memory, meanwhile avoiding cache eviction to increase the cost efficiency of the application run.
\papertitle{} is adaptive to cluster changes. In other words, a sampling phase is not required in case the cluster environment changes (e.g., new machine/instance type with larger memory size).
Also, \papertitle{} can predict the performance boundary of a resource-constrained cluster. 
It indicates if a certain cluster can efficiently run an application with higher data sizes or not.
This is particularly important for those application domains whose data sizes grow rapidly but need to pass over the same data production pipelines~\cite{Sumbaly:2013:Sigmod:LinkedInBigDataEcosystem, Migliorini:2019:arXiv:MLPipelineForPhysics, Donratanapat:2020:EMS:Flooding}.

Although an autonomous selection of cluster size for caching crucial datasets is fundamental, we are not aware of a fast and cost-effective sampling-based approach that fully addresses it.
Currently, optimal resource provisioning based on accurate prediction of the size of cached datasets remains an open challenge.
In summary, we make the following contributions: 

\begin{itemize}[itemsep=0pt,topsep=0pt]
    
	\item We introduce an efficient approach for minimizing the cost of sample runs.
	
	\item We present \papertitle{}, a lightweight sampling-based framework that predicts the size of cached datasets and selects an optimal cluster size (area C) for caching them.
	
	\item We perform extensive analysis of different machine learning applications and stress their minimal sampling requirements for an optimal cluster size selection.

\end{itemize}

\noindent

\noindent
We evaluate \papertitle{} on 8 real-world applications in the \hibench{} benchmark~\cite{Huang:2010:ICDEW:HiBench,url:HiBench}.
Relying on tiny sample datasets (\SI{0.1}{\percent} - \SI{0.3}{\percent} of the original data), \papertitle{} selects the optimal cluster size for all 8 actual runs which reduces execution cost to \SI{52.6}{\percent} compared to the average cost across all cluster sizes.
On average, \papertitle{}'s sample runs cost \SI{4.6}{\percent} of the optimal application actual run.
Using the same sample runs, we evaluate the scalability of \papertitle{} with larger data scales (\SI{150}{\percent} - \SI{18e4}{\percent} of the original data).
With these larger data scales, \papertitle{} selects the optimal cluster size in 7 out of the 8 evaluated cases.
And \papertitle{}'s sample runs cost, on average, \SI{1.08}{\percent} of the optimal application actual run.
Finally, \papertitle{} predicts for each application the maximum data scale that can be efficiently processed on a resource-constrained cluster, with less than \SI{5}{\percent} of an error.

\section{Related work}
\label{sec:related-work}



Many contributions have tried to observe, analyze, predict and optimize the execution time and cost of big data applications running on distributed systems. 
We group the following related work according to the topic they have addressed.

\textbf{Caching decision support tools} help application developers to determine which datasets shall be cached and when to purge from memory \cite{SparkCAD,li2020detecting}.
However, these tools does not consider the size of datasets and the required cluster configuration that guarantees eviction-free runs.

\textbf{Cache eviction policies} and \textbf{auto-tuning of memory configuration approaches} tackle cache limitation in a best-effort manner, but with penalties caused by cache eviction.
This makes them suitable solutions if an inappropriate cluster size is selected (area A in Figure~\ref{plot:svm-intro}) or in the case of resource-constrained clusters where computation and storage resources are not extendable. 
MRD~\cite{Perez:2018:ICPP:MRD} and LRC~\cite{Yu:2017:INFOCOM:LRC, Yu:2019:TCC:LRC} are DAG-aware cache eviction policies in Spark that rank cached datasets based on their reference distance and reference count, respectively, but without considering their size.
We apply both policies for the same \textsc{svm} experiments (depicted in Figure~\ref{plot:svm-intro}) and do not realize any performance improvement. 
This is because only one dataset is cached in \textsc{svm}.
We further study the full set of \hibench{} applications and realize that most of them cache a single dataset, at most.
For those few applications that cache multiple datasets, both eviction policies make, mostly, the same decision.
MemTune~\cite{Xu:2016:IPDPS:MEMTUNE} is a memory manager that observes memory usage during application run and dynamically re-adjusts storage and execution memory regions. 
It prioritizes execution over caching to reduce GC overhead. 
RelM~\cite{kunjir2020black} introduces a safety factor to ensure error-free execution in resource-constrained clusters. 
Besides reducing the cache eviction ratio, it also considers provisioning more memory to ensure low GC overhead and improve task concurrency. 

\textbf{Runtime prediction approaches.} Ernest~\cite{Venkataraman:2016:NSDI:Ernest} is a sampling-based framework that predicts the runtime of compute-intensive long-running Spark applications.
Although it collects training data points by applying optimal experiment design~\cite{Pukelsheim:1993:OptimalDoE} on sample datasets (1\%-10\% of the original dataset) to reduce the overhead of sample runs, however, it also reduces the number of iterations during these sample runs to make their overhead tolerable. 
Reducing the number of iterations is not always practical because tuning an application parameter like the number of iterations during sample runs requires end users (or schedulers) to have knowledge of the application and its parameters, which they might lack. 
Furthermore, some iterative applications (e.g., \emph{Logistic Regression} in \hibench{}) do not take number of iterations as parameter. 
Rather, they run until a predefined condition is met \cite{Popescu:2013:PVLDB:PREDIcT}.
Ernest execution time model considers serial parts, parallel parts, and the overhead caused by increasing the cluster size but without considering cache limitations.
Masha~\cite{al:2020:DISPA:masha} is a sampling-based framework for runtime prediction of big data applications.
In addition to parallel parts, the proposed runtime model in Masha considers the execution memory and data shuffling cost between consecutive stages.
The design of sample runs presented by Masha is not generic for different applications. 
And similar to Ernest, it does not address cache limitations issues.

\textbf{Approaches for recommendation of cluster configuration} rely on sample (or historical) runs to predict (near-to-) optimal cluster configuration.
The problem space of these studies is quite huge (thousands of different cloud configurations)~\cite{Alipourfard:2017:NSDI:CherryPick, Mahgoub:2020:ATC:OPTIMUSCLOUD, Mahgoub:2019:ATC:SOPHIA, Klimovic:2018:ATC:Selecta, Klimovic:2018:OSDI:Pocket} because they consider instance type and size (i.e., machine type and cluster size, respectively) as two optimization parameters.
Therefore, these studies propose near-to-optimal configurations rather than the optimal ones. 
However, by knowing the size of cached datasets and the cache capacity of each machine (i.e., instance type), the optimal cluster size (i.e., instance size) can be predicted and, thus, the huge problem space can be reduced by only considering the machine/instance type (Azure and AWS provide 146 and 133 different instance types, respectively~\cite{Mahgoub:2020:ATC:OPTIMUSCLOUD}).
CherryPick~\cite{Alipourfard:2017:NSDI:CherryPick} aims to be accurate enough to identify poor configurations, adaptive using a black-box approach without considering the hosting framework's internals, and fast with low overhead by applying interactive searching while constructing the model and carrying out the required experiments. 
Sophia~\cite{Mahgoub:2019:ATC:SOPHIA} is a framework that (re)configures the cluster environment on the fly. 
Selecta~\cite{Klimovic:2018:ATC:Selecta} is a framework for I/O-intensive recurring workloads that considers storage type as a new dimension of cluster configuration instead of instance type and size.
Pocket~\cite{Klimovic:2018:OSDI:Pocket} minimizes execution cost for serverless analytics by selecting suitable cluster configuration (remote storage) based on the job I/O requirements.
OptimusCloud~\cite{Mahgoub:2020:ATC:OPTIMUSCLOUD} considers two dependent configuration spaces: One 
related to the workload itself (data size, cached data, \#iterations, required execution memory) and 
the other related to the cluster (instance type and size). 
None of these contributions factor in cache limitations. Moreover, they require numerous non-short-running performance-based experiments.
Juggler~\cite{juggler} considers application parameters to recommend cluster configurations with autonomous selection of datasets for caching.
But, its offline-training overhead is not tolerable and, thus, it is limited to recurring applications. 
\section{Background}
\label{sec:background}

In this section, we discuss details of 
Spark, our system use case for distributed in-memory data processing. 

\subsection{Execution model}
\label{subsec:background_Spark_Execution_Model}

Spark runs applications on multiple \emph{executors} that perform various parallel operations on partitioned data called Resilient Distributed Dataset or RDD~\cite{Zaharia:2012:RDD}. 
A class of operations called \emph{transformations} (e.g., filter, map) create new RDDs from existing ones while another class called \emph{actions} 
(e.g., count, collect) return a value to the (\emph{driver}) program after making computations on RDDs, and store results.

An application level is the highest level of computation and consists of one or more sequential \emph{jobs}, each of which is triggered by an action. 
This means there are as many jobs in an application as there are number of actions.
A job comprises of a sequence of the transformations, represented by a DAG of transformations, followed by a single action.
When a transformation is applied on an RDD, a new one is created.
The parent-child dependency between RDDs is represented in a logical plan, by way of a lineage or DAG starting from an action up to 
either the root RDDs that are cached or original data blocks from the distributed file system.
Previous studies~\cite{Wang:2015:HPCC,hernandez:2018:FGCS:using,ousterhout:2015:NSDI:making} presented Spark's execution model in detail.
\subsection{Iterative workloads}
\label{subsec:background_Iterative_workloads}
\begin{figure}
    \centering
    \includegraphics[width=1.0\columnwidth]{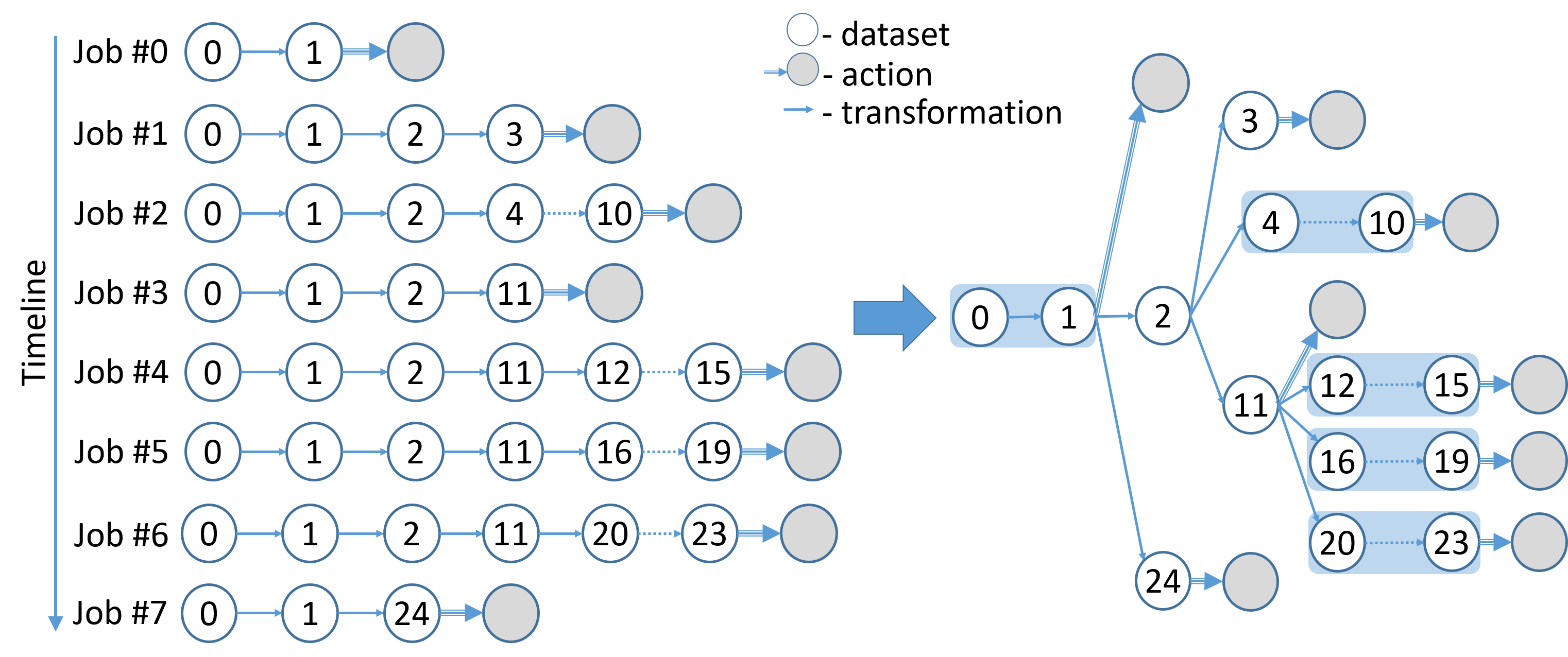}
    \vspace{-15pt}
    \caption{Merging DAGs: \emph{Logistic Regression} use case.}
    \vspace{-10pt}
    \label{fig:single_DAG_overview}
\end{figure}

As different jobs may consist of many transformations in common, we merge all their DAGs to represent an application in a single DAG of transformations, as illustrated with the \emph{Logistic Regression} application in Figure~\ref{fig:single_DAG_overview}. 
The number of times a dataset is computed is determined by the number of its child branches in the resulting DAG. 
For example, 
datasets $ \dataset{}_1 $ and $ \dataset{}_2 $ are computed 8 and 6 times respectively.
Starting from \dataset{$_0$}, computing the datasets can be traced in a depth-first traversal order.
If datasets are not cached, they will be recomputed in each iteration as follows. \dataset{$_0$} -> \dataset{$_1$} -> $action_0$ then \dataset{$_0$} 
-> \dataset{$_1$} 
-> \dataset{$_2$} -> \dataset{$_3$} -> $action_1$, and so on until \dataset{$_0$} 
-> \dataset{$_1$} 
-> \dataset{$_{24}$} -> $action_7$.
In this case, \dataset{$_0$}, \dataset{$_1$}, \dataset{$_2$}, and \dataset{$_{11}$} are recomputed 7, 7, 5, and 3 times, respectively.
In the absence of caching or in the case of cache evictions, a huge number of iterations (e.g., hundreds of iterations) increases execution time and cost significantly, as seen in Figure~\ref{plot:svm-intro}.

\subsection{Memory management}
\label{subsec:background_Memory_layout}
\begin{figure}[t!]
    \centering
    \includegraphics[width=1.0\columnwidth]{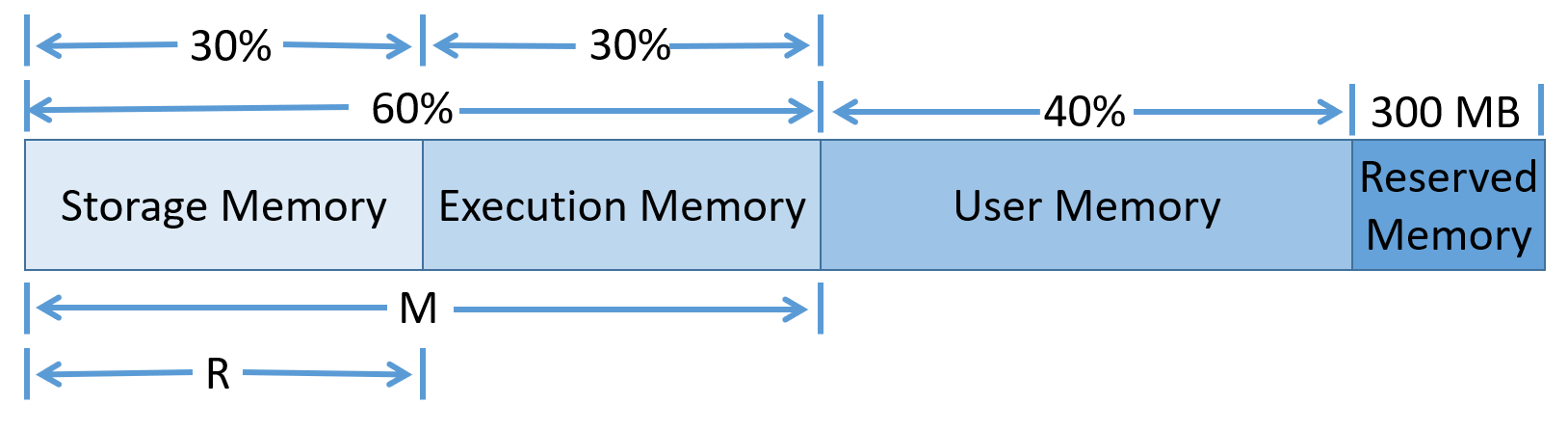}
    \caption{Spark: Memory layout.}
    \vspace{-10pt}
	\label{fig:Spark_Memory_Layout}
\end{figure}

As depicted in Figure~\ref{fig:Spark_Memory_Layout}, Spark splits memory into multiple regions. 
We focus on the storage and the execution regions, respectively used for caching datasets and computation~\cite{Zhu:2017:ICPADS:MCS}.
Both regions share the same memory space (i.e., the unified region \emph{M}) such that if the execution memory is not utilized, all the available memory space can be used for caching, and vice versa. 
There is a minimum storage space \emph{R} below which cached data is not evicted. 
That is, in each executor, at least R and at most M can be utilized to cache datasets. 
The sizes of R and M are configurable~\cite{Apache:Spark:Conf}. 
Partitions of least recently used cached RDDs are evicted when the size of cached partitions 
exceeds M or 
exceeds R while the execution memory is utilized.

\section{Efficient sample runs}
\label{sec:efficient-sampling}

In this section, we explain how to minimize the cost of sample runs with empirical evaluations. 
Specifically, we show that the sample run phase required for predicting the size of the cached datasets is less challenging than that required for execution time prediction.
As previous studies tackle data sampling challenges \cite{hamidi2018analysis, chakaravarthy2009analysis, nirkhiwale2013sampling}, we do not address them in this work.
Similar to Ernest~\cite{Venkataraman:2016:NSDI:Ernest} and Masha~\cite{al:2020:DISPA:masha}, we assume that sample datasets are available.

\subsection{Size of sample runs}
\label{subsec:sample-run-size}

Few sample runs are sufficient to predict the size of the cached datasets because the operation of data flows is deterministic generally.
For example, if we conduct two short-running experiments of the same application using the same data and same cluster configuration, the sizes of cached (and non-cached) datasets do not vary. 
However, this is not the case regarding execution time. 

\begin{figure}[t!]
\centering
\hspace*{-12.5pt} 
\begin{tikzpicture}
 \begin{axis}[
    legend style={at={(0.3,1.2)},
    draw=none,
    /tikz/every even column/.append style={column sep=0.5cm},
    anchor=north,
    legend columns=-1},
    height=125pt,
    width=1.0\columnwidth,
    axis y line*=left,
    xmin=1,
    xmax=30,
    xlabel=Data scale,
    ylabel=Execution time (sec),
    xtick=data,
    xticklabels=\empty,
    extra x ticks={5,15,25},
    extra x tick labels={1,2,3},
    extra x tick style={
    tick label style={yshift=-1mm}
    }
    ]
    \addplot[line width=0.3mm, color=blue] table[y=ExecutionTime, col sep=comma] {data/sample-run-size.dat}; 
    \legend{Time}
 \end{axis}
     
 \begin{axis}[
   legend style={at={(0.6,1.2)},
   draw=none,
   /tikz/every even column/.append style={column sep=0.5cm},
   anchor=north,
   legend columns=-1},
   height=125pt,
   width=1.0\columnwidth,
   axis y line*=right,
   xmin=1,
   xmax=30,
   ylabel=Size of cached dataset (MB),
   xtick=data,
   xticklabels=\empty,
   ]
   \addplot[line width=0.3mm, color=red] table[y=IntermediateResultsSize, col sep=comma] {data/sample-run-size.dat}; 
   \legend{Size}
 \end{axis}

\draw (0,0) -- (0,-0.5);
\draw (2.14,0) -- (2.14,-0.5);
\draw (4.515,0) -- (4.515,-0.5);
\draw (6.89,0) -- (6.89,-0.5);

\end{tikzpicture}
\vspace{-20pt}
\caption{Short-running experiments on 3 data scales (\textsc{svm}).}
\vspace{-15pt}
\label{fig:30runs}
\end{figure}
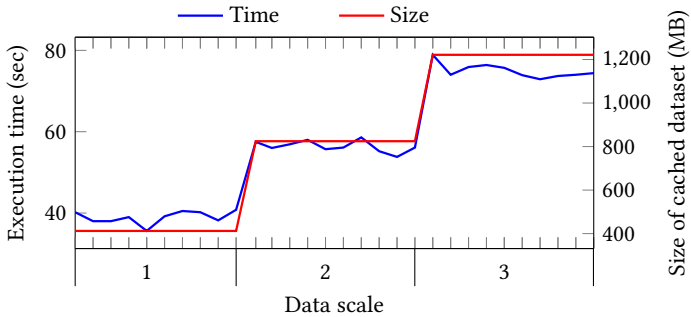

To validate this, we select \textsc{svm}, which caches one dataset, to run 10 experiments on \SI{738.1}{\mega\byte} (data scale 1, 12 blocks), 10 experiments on \SI{1501.6}{\mega\byte} (data scale 2, 24 blocks) and 10 experiments on \SI{2.2}{\giga\byte} (data scale 3, 36 blocks). 
We conduct all runs on a single machine. 
As illustrated in Figure~\ref{fig:30runs}, we see that the size of the cached dataset remains constant in all runs of the same data scale. 
Also, we notice a considerable variance in execution time between the runs of the same data scale, which affects the construction and training of prediction models. 
One way to overcome this problem is to run several experiments on the same data scale and obtain the statistical average (or median). 
Another way 
is to increase the size of sample datasets to make sample runs longer and, thus, the execution time variance relatively lower. 
However, both solutions increase the cost of sample runs tremendously, which explains why runtime prediction approaches are limited to long-running applications. 

To build robust models for predicting the size of the cached datasets in big data scale, we carry out sample runs on tiny datasets within the range of \SIrange{0.1}{0.3}{\percent} of the original data.

\subsection{Parallelism}
\label{subsec:paralellism}

Distributed file systems (e.g., HDFS) store original data by fragmenting it into equal chunks, namely blocks.
The size of blocks are configurable~\cite{Apache:HadoopConf} (64 or 128 MB by default).
In order to decrease the data size during sample runs, we either 
\begin{inparaenum}[(1)]
 \item reduce the size of each block (\bsize{}), or 
 \item select few data blocks (\bnumber{}).
\end{inparaenum}
For example, if the block size is configured to be 64 MB, 1 TB of data is stored in 16K blocks.
Thus, 16 blocks out of them could be selected for a sample run of \SI{0.1}{\percent} of the original data.

\bnumber{} is less costly than \bsize{} because it only requires selecting data blocks from a distributed file system during sample runs, whereas \bsize{} 
brings extra overhead in preparing the sample data. 
And since we are not expecting memory limitation during sample runs, increasing the parallelism increases the execution time of each sample run (i.e., data shuffling and cleaning). 

In order to validate this, we conduct two runs of \textsc{svm} with an input data of \SI{1.2}{\giga\byte} on a single machine. 
The number of data blocks (i.e., number of tasks) in the first run is 10 and it takes \SI{41}{\second}.
In the second run, the number of data blocks is 1000 and it takes \SI{3.5}{\minute}. 
In addition, during the first and second run, the size of the cached dataset is \SI{728.9}{\mega\byte} and \SI{747.8}{\mega\byte}, respectively. 
This shows that the size of datasets is influenced by the parallelism level.
Hence, in the case of \bnumber{}, 
if we significantly reduce the number of tasks during sample runs, then predicting the size of the cached datasets might be affected. 
To tackle this problem, we always keep the number of tasks proportional to the data scale by fixing the block size. 
For example, if the full-scale dataset cosists of 16K blocks, then the sample runs with \SI{0.1}{\percent}, \SI{0.2}{\percent} and \SI{0.3}{\percent} of the input data scale will contain 16, 32, and 48 tasks respectively.

For some compute-intensive applications, the size of the original data is relatively small (as we will show in Section~\ref{sec:evaluation}) and, thus, the number of its blocks in the distributed file system is not enough to apply \bnumber{}. 
In such cases, \bsize{} is used in spite of its costs.
\subsection{Cluster configuration}
\label{subsec:cluster-configuration}

We carry out all sample runs on a single machine to reduce the cost of sample runs. 
The serial part of a short-running experiment is relatively high compared with the parallel part and, hence, adding more machines during a sample run might not speed up the execution time. 
Rather, it leads to higher execution cost because of the increased overhead of negotiating resources (e.g., by \textsc{yarn}) and the increase in data transfer overhead with addition of more machines. 
\begin{figure*}[t!]
    \centering
    \includegraphics[width=1.0\textwidth]{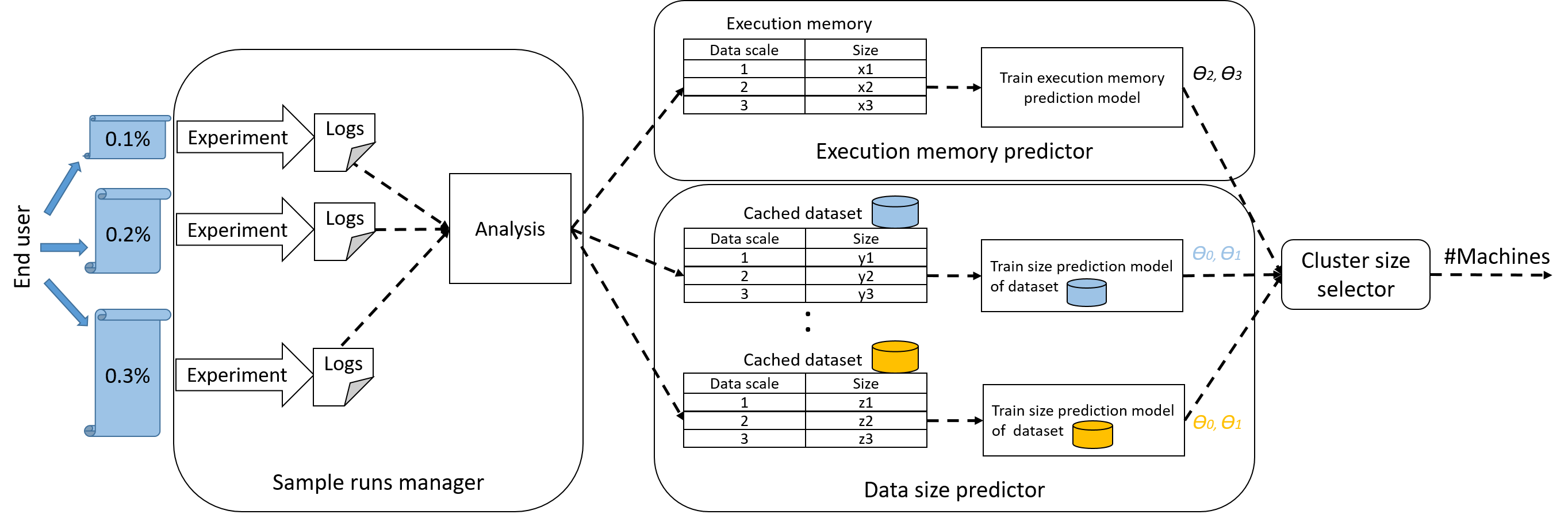}
    \vspace{-15pt}
    \caption{Overview of \papertitle{}.}
    \vspace{-10pt}
    \label{fig:blink_overview_detailed}
\end{figure*}
To validate this, we run \textsc{svm} on \SI{1.2}{\giga\byte} input data using a single machine and also using 12 machines. 
The execution cost on 12 machines is $ 13.9\times $ higher than on a single machine. 
The exception that makes carrying out sample runs on a single machine too costly is when cached datasets do not fit in memory of a single machine.
However, this is unlikely for sample runs with tiny datasets.

\subsection{Number of sample runs}
\label{subsec:number-of-sample-runs}

Our experiments with all applications in \hibench{} show that the prediction models for the size of the cached (and non-cached) datasets with respect to the input data scale are linear. 
Therefore, two sample runs are sufficient to construct a model. 
However, knowing that sample runs are lightweight, more sample runs could be conducted to apply \emph{cross validation} to choose a well-fitting 
model.

\section{Blink}
\label{sec:blink}

In this section, we present \papertitle{}, a lightweight sampling-based framework that performs autonomous and optimal resource provisioning for iterative big data applications.
As depicted in Figure~\ref{fig:blink_overview_detailed}, firstly, \textit{Sample runs manager}~(Section~\ref{subsec:light-weight-sampling}) carries out lightweight sample runs on \SIrange{0.1}{0.3}{\percent} data samples of the original data. 
Then, based on these sample runs, \textit{size predictor}~(Section~\ref{subsec:data-size-prediction}) and \textit{execution memory predictor}~(Section~\ref{subsec:execution-memory-prediction}) train prediction models to predict the size of cached datasets and the required amount of execution memory per machine in the actual run, respectively. 
Finally, based on the extracted models and caching capacity of each machine, \textit{cluster size selector}~(Section~\ref{subsec:cluster-configuration-selection}) selects the optimal cluster size that guarantees eviction-free actual runs.

\subsection{Sample runs manager}
\label{subsec:light-weight-sampling}

Sample runs manager carries out three sample runs on tiny data samples (\SIrange{0.1}{0.3}{\percent} of the original data).
It carries out the sample runs on a single machine and monitors every single run to make quick decisions regarding the following atypical cases:
\begin{itemize}[itemsep=0pt,topsep=0pt]
    \item If there is no cached dataset in the application, sample runs manager directly selects a single machine for the actual runs (i.e., the longest execution time but the cheapest cost).
    \item If there is a cached dataset and eviction occurs, which is unusual while handling tiny datasets, it terminates the sample run and carries out new ones with lower sampling scales.
\end{itemize}

\noindent 
While conducting sample runs, \emph{SparkListener} collects runtime metrics and stores them as log files in the distributed file system (e.g., HDFS). 
Sample runs manager analyzes the logs and collects the size of each cached dataset.
\subsection{Data size predictor}
\label{subsec:data-size-prediction}

After carrying out sample runs, the data size predictor trains a set of models to predict the size of cached datasets in actual runs. 
For each cached dataset, the data size predictor takes the scale of the data sample as a feature and its size as a label.
Thus, the scales in sample runs are 1, 2, and 3; while in the actual run, the scale is 1000. 
The data size predictor applies \emph{cross validation} to determine the error of each model. 
It does this by keeping each point among the three training experiments, in turn, as a test experiment and fitting the model with the remaining 2 experiments. 
Our experiments show that the sizes of all cached datasets fit into the model below, although the data size predictor evaluates many other models:
\begin{equation} \label{eq:size-prediction}
     \dataset{}_{size} = \theta_{0} + \theta_{1} \times datascale
\end{equation}
\noindent
We use the \emph{curve\_fit} solver~\cite{url:CURVE} with enforced positive bounds to train the models while avoiding negative coefficients, and Root Mean Square Error (RMSE) to evaluate the models.
\subsection{Execution memory predictor}
\label{subsec:execution-memory-prediction}

The minimum and the maximum amount of memory for caching in each machine can be known (M and R in~Figure~\ref{fig:Spark_Memory_Layout}) and, in turn, the minimum and the maximum number of machines can be determined using the following equations:
\begin{gather*}
    Machines_{min} = {\lceil}\frac{\sum_{}^{CachedDs} \dataset{}_{size}}{M} {\rceil}\\
    Machines_{max} = {\lceil}\frac{\sum_{}^{CachedDs} \dataset{}_{size}}{R} {\rceil} 
\end{gather*}
where $\sum_{}^{CachedDs} \dataset{}_{size}$ is the total size of cached datasets, R is the memory region used for caching and M is the unified memory region for both caching and execution (cf.~Figure~\ref{fig:Spark_Memory_Layout}). 
Selecting less than $Machines_{min}$ leads to cache eviction because utilizing the whole unified memory space (i.e., M) in each machine for caching will not be enough to cache all datasets. 
In contrast, allocating more than $Machines_{max}$ gives no caching benefits since utilizing the storage memory (i.e., R) in each machine will be enough for caching all datasets.
In other words, $ Machines_{max} $ is required to cache datasets without eviction, when the entire $($M$-$R$)$ memory region is utilized for execution. 
If M is not utilized at all, then the entire region can be used for caching and, hence, $ Machines_{min} $ is required to cache datasets without evictions. 
Considering that the gap between $ Machines_{min} $ and $ Machines_{max} $ may be quite wide and the execution memory utilization differs from one application to another, there is a need for a precise prediction 
of the amount of memory required for execution. 
Similar to the data size predictor (cf.~Section~\ref{subsec:data-size-prediction}), the execution memory predictor analyzes the execution memory usage in sample runs and trains linear models to predict the total amount of execution memory required for the actual runs. 
Our experiments show that the relationship between the data sample scale and the amount of execution memory fits into the following model, although the execution memory predictor evaluates many other models:
\begin{align*}
    Memory_{execution} &= \theta_{2} + \theta_{3} \times datascale
\end{align*}

\subsection{Cluster size selector}
\label{subsec:cluster-configuration-selection}
\noindent
Based on M and R in Figure~\ref{fig:Spark_Memory_Layout} (which are derived from machine/instance type), the cluster size selector calculates the required amount of memory for execution in each machine as follows:
\begin{equation*} \label{eq:machine-execution-memory}
     MachineMemory_{execution} = \min (M-R, \frac{Memory_{execution}}{Machines}) 
\end{equation*}
Then, it selects the minimal number of machines that fulfills the following condition:
\begin{equation*} \label{eq:machine-number-selection}
    \frac{\sum_{}^{CachedDs} \dataset{}_{size}}{Machines} < (M-MachineMemory_{execution}) \times Machines
\end{equation*}
Note that \papertitle{} constructs the prediction models only once, and then reuses them to predict the optimal sizes for various clusters with different machine types.
In addition, in multi-tenant environments, the recommended cluster configuration is not affected by concurrent application runs hosted on the same machines because they are deployed in isolated virtual machines, and cluster managers (e.g., YARN~\cite{Vavilapalli:2013:AHY:2523616.2523633}) do not offer an occupied memory region (i.e., M) to newly submitted applications. 

\section{Evaluation}
\label{sec:evaluation}
For evaluation, we use 8 iterative, real-world machine learning applications from \hibench{}: \emph{Alternating Least Squares} (\textsc{als}), \emph{Bayesian Classification} (\textsc{bayes}), \emph{Gradient Boosted Trees} (\textsc{gbt}), \emph{K-means clustering} (\textsc{km}), \emph{Logistic Regression} (\textsc{lr}), \emph{Principal Components Analysis} (\textsc{pca}), \emph{Random Forest Classifier} (\textsc{rfc}), and \emph{Support Vector Machine} (\textsc{svm}). 
We exclude applications in \hibench{} that do not cache any dataset in memory.  

\textbf{Sample runs.}
For conducting sample runs and measuring the robustness of the extracted models for re-usability on clusters with different machine types, we use a single node -- 
Intel Core i3-2370M CPU running at 4 x 2.40GHz, 3.8 GB DDR3 RAM, and 388 GB disk. 
For each application, we run 3 lightweight runs on sample data size in the range of \SI{0.1}{\percent} - \SI{0.3}{\percent} of the complete input data scale.

\textbf{Actual runs.}
We made all actual runs on a private 12-node
cluster equipped with Intel Core i5 CPU running at 4x 2.90 GHz, 16 GB DDR3 RAM, 1 TB disk, and 1 GBit/s LAN. All nodes (including the previously mentioned single node used for sample runs) used in the experiments run Hadoop MapReduce 2.7, Spark 2.4.0, Java 8u102, and Apache Yarn on top of HDFS.

To evaluate whether \papertitle{} recommends the optimal cluster size or not, we run each application on all cluster sizes (from 1 to 12 machines), as shown in Table~\ref{tbl:application-details} (with data scale \SI{100}{\percent}). 
We applied \bnumber{} sampling approach, on \textsc{bayes}, \textsc{lr}, \textsc{rfc}, and \textsc{svm}; and applied \bsize{} sampling approach, on \textsc{als}, \textsc{gbt}, \textsc{km}, \textsc{pca}.

\begin{table*}[tb]
\centering
\small{
\npdecimalsign{.}
\begin{adjustbox}{max width=\textwidth}
\begin{tabular}{ c c r r r r r r r r r r r r r r r r}
\toprule
\multirow{2}*{\if 0 \textbf{Phase} \fi} &\multirow{2}*{\textbf{\#Machines}} & \multicolumn{2}{c}{\textsc{als}} &\multicolumn{2}{c}{\textsc{bayes}} &\multicolumn{2}{c}{\textsc{gbt}} &\multicolumn{2}{c}{\textsc{km}} &\multicolumn{2}{c}{\textsc{lr}} &\multicolumn{2}{c}{\textsc{pca}} &\multicolumn{2}{c}{\textsc{rfc}} &\multicolumn{2}{c}{\textsc{svm}} \\
\cmidrule(lr){3-4}\cmidrule(lr){5-6}\cmidrule(lr){7-8}\cmidrule(lr){9-10}\cmidrule(lr){11-12}\cmidrule(lr){13-14}\cmidrule(lr){15-16}\cmidrule(lr){17-18}
 &  & Time & Cost & Time & Cost & Time & Cost & Time & Cost & Time & Cost & Time & Cost & Time & Cost & Time & Cost \\

\midrule
Sample runs & 1 & 5.8 & 5.8 & 1.4 & 1.4 & 1.4 & 1.4 & 1.2 & 1.2 & 1.0 & 1.0 & 7.7 & 7.7 & 3.9 & 3.9 & 1.2 & 1.2 \\
 \cmidrule(lr){1-18}
 Approach &  &  \multicolumn{2}{c}{\bsize{}}  &  \multicolumn{2}{c}{\bnumber{}}  &  \multicolumn{2}{c}{\bsize{}}  &  \multicolumn{2}{c}{\bsize{}}  &  \multicolumn{2}{c}{\bnumber{}}  &  \multicolumn{2}{c}{\bsize{}}  &  \multicolumn{2}{c}{\bnumber{}}  &  \multicolumn{2}{c}{\bnumber{}}  \\
\cmidrule(lr){1-18}
 Scale \SI{100}{\percent} (size) &  &  \multicolumn{2}{c}{5.6 GB}  &  \multicolumn{2}{c}{17.6 GB}  &  \multicolumn{2}{c}{30.6 MB}  &  \multicolumn{2}{c}{21.5 GB}  &  \multicolumn{2}{c}{22.4 GB}  &  \multicolumn{2}{c}{1.5 GB}  &  \multicolumn{2}{c}{29.8 GB}  &  \multicolumn{2}{c}{59.6 GB}  \\
\cmidrule(lr){1-18}
 Scale \SI{100}{\percent} (\#Blocks) &  & \multicolumn{2}{c}{100}  & \multicolumn{2}{c}{2K} & \multicolumn{2}{c}{100} & \multicolumn{2}{c}{200} & \multicolumn{2}{c}{2K} &  \multicolumn{2}{c}{50} & \multicolumn{2}{c}{2K}  & \multicolumn{2}{c}{2K} \\
\cmidrule(lr){1-18}
 & 1 & \cellcolor{green!100}\textbf{27.2} & \cellcolor{green!100}\textbf{27.2} & 63.3 & 63.3 & \cellcolor{green!100}\textbf{9.8} & \cellcolor{green!100}\textbf{9.8} & 137.2 & 137.2 & 337 & 337 & \cellcolor{green!100}\textbf{77.4} & \cellcolor{green!100}\textbf{77.4} & 361.6 & 361.6 & 804.8 & 804.8 \\
  & 2 & \cellcolor{green!100}14.5 & \cellcolor{green!100}29.0 & 29.1 & 58.2 & \cellcolor{green!100}6.3 & \cellcolor{green!100}12.6 & 45.4 & 90.9 & 133.5 & 266.9 & \cellcolor{green!100}41.9	& \cellcolor{green!100}83.9 & 125.4 & 250.7 & 325.6 & 651.2 \\
 & 3 & \cellcolor{green!100}9.6	& \cellcolor{green!100}28.8 & 22.2 & 66.5 & \cellcolor{green!100}5.2 & \cellcolor{green!100}15.6 & 18.2 & 54.5 & 47.6 & 142.7 & \cellcolor{green!100}30.7 & \cellcolor{green!100}92 & 91.2 & 273.6 & 172.3 & 516.9 \\
 & 4 & \cellcolor{green!100}8.7 & \cellcolor{green!100}34.9 & 14.3 & 57.1 & \cellcolor{green!100}8.7 & \cellcolor{green!100}34.9 & \cellcolor{green!100}\textbf{3.5} & \cellcolor{green!100}\textbf{13.9} & 17.3 & 69.3 & \cellcolor{green!100}28.8 & \cellcolor{green!100}115.3 & \cellcolor{green!100}\textbf{60.3} & \cellcolor{green!100}\textbf{241} & 88.5 & 354.1 \\
 & 5 & \cellcolor{green!100}8.3	& \cellcolor{green!100}41.4 & 11 & 54.8 & \cellcolor{green!100}6.9 & \cellcolor{green!100}34.5 & \cellcolor{green!100}3.2 & \cellcolor{green!100}15.8 & \cellcolor{green!100}\textbf{8.6} & \cellcolor{green!100}\textbf{42.9} & \cellcolor{green!100}26.7 & \cellcolor{green!100}133.3 & \cellcolor{green!100}52.3 & \cellcolor{green!100}261.3 & 40.7 & 203.3 \\
Actual runs & 6 & \cellcolor{green!100}7.5 & \cellcolor{green!100}45.2 & 10.1 & 60.8 & \cellcolor{green!100}5 & \cellcolor{green!100}29.9 & \cellcolor{green!100}2.7 & \cellcolor{green!100}16.5 & \cellcolor{green!100}7.7 & \cellcolor{green!100}46 & \cellcolor{green!100}25.2 & \cellcolor{green!100}151.2 & \cellcolor{green!100}51.4 & \cellcolor{green!100}308.4 & 15.7 & 94.4 \\
 (100\% data scale) & 7 & \cellcolor{green!100}4.5 & \cellcolor{green!100}31.4 & \cellcolor{green!100}\textbf{4.1} & \cellcolor{green!100}\textbf{28.5} & \cellcolor{green!100}7.7 & \cellcolor{green!100}53.9 & \cellcolor{green!100}2.1 & \cellcolor{green!100}14.8 & \cellcolor{green!100}7.2 & \cellcolor{green!100}50.6 & \cellcolor{green!100}24.8 & \cellcolor{green!100}173.3 & \cellcolor{green!100}46.5 & \cellcolor{green!100}325.4 & \cellcolor{green!100}\textbf{9.6} & \cellcolor{green!100}\textbf{67.2} \\
 & 8 & \cellcolor{green!100}4.1 & \cellcolor{green!100}33.1 & \cellcolor{green!100}3.8 & \cellcolor{green!100}30.4 & \cellcolor{green!100}4 & \cellcolor{green!100}32.2 & \cellcolor{green!100}2.3 & \cellcolor{green!100}18.8 & \cellcolor{green!100}6.9 & \cellcolor{green!100}55.6 & \cellcolor{green!100}22.4 & \cellcolor{green!100}179.5 & \cellcolor{green!100}47.2 & \cellcolor{green!100}377.8 & \cellcolor{green!100}8.6 & \cellcolor{green!100}68.9 \\
 & 9 & \cellcolor{green!100}3.9 & \cellcolor{green!100}35.2 & \cellcolor{green!100}3.7 & \cellcolor{green!100}33.2 & \cellcolor{green!100}4.7 & \cellcolor{green!100}42 & \cellcolor{green!100}2.1 & \cellcolor{green!100}18.9 & \cellcolor{green!100}6.4 & \cellcolor{green!100}57.6 & \cellcolor{green!100}20.9 & \cellcolor{green!100}187.9 & \cellcolor{green!100}41.2 & \cellcolor{green!100}370.5 & \cellcolor{green!100}8.4 & \cellcolor{green!100}75.2 \\
 & 10 & \cellcolor{green!100}3.6 & \cellcolor{green!100}36.3 & \cellcolor{green!100}3.5 & \cellcolor{green!100}35.3 & \cellcolor{green!100}6.2 & \cellcolor{green!100}62 & \cellcolor{green!100}1.9 & \cellcolor{green!100}19.3 & \cellcolor{green!100}6.3 & \cellcolor{green!100}63 & \cellcolor{green!100}19.5 & \cellcolor{green!100}194.6 & \cellcolor{green!100}39.8 & \cellcolor{green!100}397.5 & \cellcolor{green!100}8.3 & \cellcolor{green!100}83.5 \\
 & 11 & \cellcolor{green!100}3.6 & \cellcolor{green!100}39.6 & \cellcolor{green!100}3.5 & \cellcolor{green!100}38.3 & \cellcolor{green!100}5.5 & \cellcolor{green!100}60.6 & \cellcolor{green!100}1.9 & \cellcolor{green!100}21.4 & \cellcolor{green!100}5.9 & \cellcolor{green!100}65.2 & \cellcolor{green!100}18.6 & \cellcolor{green!100}204.4 & \cellcolor{green!100}40.2 & \cellcolor{green!100}442.3 & \cellcolor{green!100}8.4 & \cellcolor{green!100}92.5 \\
 & 12 & \cellcolor{green!100}3.2 & \cellcolor{green!100}38.9 & \cellcolor{green!100}3.4 & \cellcolor{green!100}41 & \cellcolor{green!100}6.1 & \cellcolor{green!100}72.9 & \cellcolor{green!100}1.9 & \cellcolor{green!100}23.2 & \cellcolor{green!100}5.5 & \cellcolor{green!100}66.2 & \cellcolor{green!100}18.3 & \cellcolor{green!100}219.1 & \cellcolor{green!100}36.7 & \cellcolor{green!100}440.6 & \cellcolor{green!100}7.7 & \cellcolor{green!100}92.9 \\
 & Avg & 8.2 & 35.1 & 14.3 & 47.3 & 6.3 & 38.4 & 18.5 & 37.1 & 49.2 & 105.2 & 29.6 & 151 & 82.8 & 337.6 & 124.9 & 258.7 \\
 
\cmidrule(lr){1-18}

Scale +\SI{150}{\percent} (size) &  &  \multicolumn{2}{c}{\SI{e3}{\percent} (56.0 GB)}  & \multicolumn{2}{c}{\SI{150}{\percent} (26.4 GB)} &  \multicolumn{2}{c}{\SI{18e4}{\percent} (53.7 GB)} & \multicolumn{2}{c}{\SI{200}{\percent} (43.0 GB)}  &  \multicolumn{2}{c}{\SI{200}{\percent} (44.7 GB)}  &  \multicolumn{2}{c}{\SI{5e3}{\percent} (74.8 GB)}  &  \multicolumn{2}{c}{\SI{200}{\percent} (59.6 GB)} &  \multicolumn{2}{c}{\SI{150}{\percent} (89.4 GB)} \\
\cmidrule(lr){1-18}
 & 5 & x & x & 20.9 & 104.4 & 1545.1 & 7725.5 & 39.2 & 195.8 & 110.9 & 554.5 & 839.4 & 4197.0 & 120.8 & 604.1 & 141.3 & 706.5 \\
 & 6 & x & x & 16.5 & 99.1 & 243.8 & 1462.8 & 22.1 & 132.5 & 80.0 & 479.9 & 689.0 & 4134.2 & 98.5 & 591.3 & 111.4 & 668.4 \\
 & 7 & x & x & 13.7 & 96.2 & \cellcolor{green!100}\textbf{174.3} & \cellcolor{green!100}\textbf{1220.1} & \textbf{14.9}  & \textbf{104.3} & 63.3 & 442.9 & \cellcolor{green!100}\textbf{558.3} & \cellcolor{green!100}\textbf{3908.0} & \cellcolor{green!100}\textbf{80.9} & \cellcolor{green!100}\textbf{566.3} & 91.5 & 640.4 \\
Actual runs & 8 & x & x & 11.0 & 88.2 & \cellcolor{green!100}167.1 & \cellcolor{green!100}1336.8 & \cellcolor{green!100}4.6 & \cellcolor{green!100}36.4 & 43.7 & 350.0 & \cellcolor{green!100}511.0 & \cellcolor{green!100}4088.3 & \cellcolor{green!100}78.6 & \cellcolor{green!100}628.9 & 44.9 & 389.0 \\
 (+150\% data scale) & 9 & \cellcolor{green!100}\textbf{123.5} & \cellcolor{green!100}\textbf{1111.2} & 9.4 & 84.3 & \cellcolor{green!100}170.5 & \cellcolor{green!100}1534.5 & \cellcolor{green!100}4.3 & \cellcolor{green!100}39 & 24.6 & 221.5 & \cellcolor{green!100}528.4 & \cellcolor{green!100}4756.0 & \cellcolor{green!100}74.6 & \cellcolor{green!100}671.4 & 19.9 & 178.8 \\
 & 10 & \cellcolor{green!100}154.9 & \cellcolor{green!100}1549.4 & 7.3 & 72.6 & \cellcolor{green!100}150.4 & \cellcolor{green!100}1504.4 & \cellcolor{green!100}3.2 & \cellcolor{green!100}31.8 & 22.0 & 220.5 & \cellcolor{green!100}473.3 & \cellcolor{green!100}4733.4 & \cellcolor{green!100}71.8 & \cellcolor{green!100}718 & \cellcolor{green!100}\textbf{16.0} & \cellcolor{green!100}\textbf{159.8} \\
 & 11 & \cellcolor{green!100}107.6 & \cellcolor{green!100}1138.5 & \cellcolor{green!100}\textbf{5.1} & \cellcolor{green!100}\textbf{56.1} & \cellcolor{green!100}137.4 & \cellcolor{green!100}1511.4 & \cellcolor{green!100}3.2 & \cellcolor{green!100}35.2 & 16.2 & 178.0 & \cellcolor{green!100}483.6 & \cellcolor{green!100}5319.4 & \cellcolor{green!100}76.7 & \cellcolor{green!100}843.9 & \cellcolor{green!100}15.0 & \cellcolor{green!100}165.2 \\
 & 12 & \cellcolor{green!100}116.1 & \cellcolor{green!100}1393.2 & \cellcolor{green!100}5.1 & \cellcolor{green!100}61.6 & \cellcolor{green!100}121.4 & \cellcolor{green!100}1456.8 & \cellcolor{green!100}3.1 & \cellcolor{green!100}36.9 & \cellcolor{green!100}\textbf{14.7} & \cellcolor{green!100}\textbf{176.3} & \cellcolor{green!100}432.8 & \cellcolor{green!100}5193.7 & \cellcolor{green!100}76.6 & \cellcolor{green!100}919.3 & \cellcolor{green!100}13.8 & \cellcolor{green!100}166.2 \\
\bottomrule
\end{tabular}
\end{adjustbox}
\vspace{-10pt}
\caption{Overview of evaluated \hibench{} applications. Recommended cluster size is shown in bold.\\ Colored cells refer to cluster sizes that do not cause cache evictions. \\ x refers to failures caused by memory limitation. \\ Time unit is represented in minutes. Cost unit is represented in machine minutes.}
\label{tbl:application-details}
\vspace{-10pt}
}
\end{table*}


\subsection{Selected cluster size}
\label{subsec:optimal-configuration-selection}

As mentioned in Section~\ref{sec:introduction}, we consider an optimal cluster size as the minimum number of machines that fit all cached datasets in memory without cache eviction.
The green-colored cells in Table~\ref{tbl:application-details} show the cluster sizes where no eviction occurred, while the bold numbers indicate the cluster sizes selected by \papertitle{} for each application. 
Table~\ref{tbl:application-details} (with data scale \SI{100}{\percent}) shows that for all applications, \papertitle{} selects the optimal cluster size.
We can observe this by a comparison between the first green-colored cell and the bold number for each application actual run.

\begin{figure}[t!]
\centering
\begin{tikzpicture}
\begin{axis}[ybar,
             bar width=4.5pt,
             xtick=data,
             ymode = log,
             ymin=0,
             ymax=1000,
             ytick={12, 25, 50, 100, 200, 400, 800},
             yticklabels={12, 25, 50, 100, 200, 400, 800},
             ylabel style={align=center},
             ylabel=Execution cost\\(Machine min),
             width=1.0\columnwidth,
             legend style={at={(0.5,1.5)},
             draw=none,
             /tikz/every even column/.append style={column sep=0.5cm},
             anchor=north,
             legend columns=-1},
             enlarge x limits = 0.05,
             enlarge y limits = 0.005,
             height=100pt,
             ymajorgrids,
             table/col sep=comma,
             table/x=Application,
             xtick={0,...,8},
             cycle list/Blues-3,
             every axis plot/.append style={fill},
             xticklabels={{\textsc{als}},{\textsc{bayes}},{\textsc{gbt}},{\textsc{km}},{\textsc{lr}},{\textsc{pca}},{\textsc{rfc}},{\textsc{svm}}},
             xtick style={draw=none},
             ytick style={draw=none},
             legend style={at={(0.5,1.31)},
             draw=none,
             /tikz/every even column/.append style={column sep=0.5cm},
             anchor=north,
             legend columns=-1},
             ]
\addplot[draw=none, fill=Blues-D] table [x expr=\coordindex, y=BlinkCost] {data/cost-optimization.dat};
\label{tikz:BlinkCost}
\addplot[draw=none, fill=Blues-F] table [x expr=\coordindex, y=AverageCost] {data/cost-optimization.dat};
\label{tikz:AverageCost}
\addplot[draw=none, fill=Blues-I] table [x expr=\coordindex, y=WorstCost] {data/cost-optimization.dat};
\label{tikz:WorstCost}

\legend{Blink, Average, Worst}
\end{axis}
\end{tikzpicture}
\vspace{-20pt}
\caption{\papertitle{} cost optimization.}
\vspace{-15pt}
\label{fig:blink-cost}
\end{figure}
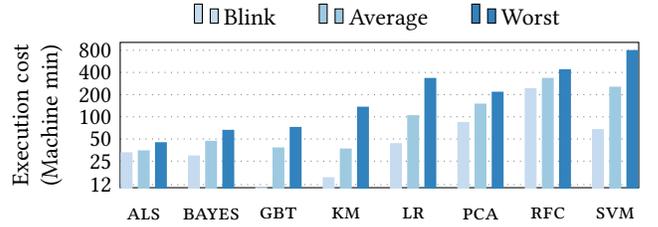

To evaluate the efficiency of \papertitle{}, we compare the sum of sample runs cost and actual run cost for the cluster size selected by \papertitle{} to the average and worst costs of actual runs. 
Figure~\ref{fig:blink-cost} shows that compared to the average and the worst costs, \papertitle{} reduces the cost to \SI{52.6}{\percent} and \SI{25.1}{\percent}, respectively. 
In some cases, the worst cluster size (that leads to the highest cost) is a single machine due to lots of recomputations (e.g., \textsc{svm}) and in other cases, it is the maximum cluster size because resources are wasted during data shuffling and processing of serial parts (e.g., \textsc{rfc}).

\subsection{Prediction accuracy}

\label{subsec:prediction-accuracy-of-cached-intermediate-results}

\begin{figure}[t!]
\centering
\begin{tikzpicture}
\begin{axis}[ybar,
             bar width=10pt,
             xtick=data,
             ymin=0,
             ylabel=Error (\%),
             width=1.0\columnwidth,
             enlarge x limits = 0.05,
             enlarge y limits = 0.005,
             height=100pt,
             ymajorgrids,
             table/col sep=comma,
             table/x=Application,
             xtick={0,...,8},
             cycle list/Blues-3,
             every axis plot/.append style={fill},
             xticklabels={{\textsc{als}},{\textsc{bayes}},{\textsc{gbt}},{\textsc{km}},{\textsc{lr}},{\textsc{pca}},{\textsc{rfc}},{\textsc{svm}}},
             xtick style={draw=none},
             ytick style={draw=none},
             legend style={at={(0.5,1.5)},
             draw=none,
             /tikz/every even column/.append style={column sep=0.5cm},
             anchor=north,
             legend columns=-1},
             ]
\addplot[draw=none, fill=Blues-I] table [x expr=\coordindex, y=Error] {data/prediction-error.dat};
\label{pgfplots:prediction-error}
\end{axis}
\end{tikzpicture}
\vspace{-5pt}
\caption{\papertitle{} prediction error of cached dataset sizes.}
\vspace{-15pt}
\label{fig:size-prediction}
\end{figure}
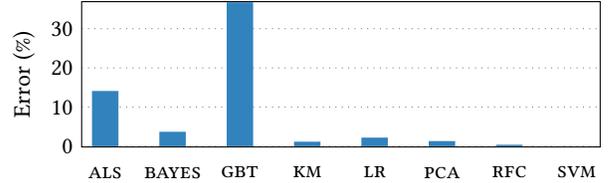

We compare the size of cached datasets in actual runs (cf.~Table~\ref{tbl:application-details} with data scale \SI{100}{\percent}) with the ones that data size predictor (cf.~Section~\ref{subsec:data-size-prediction}) predicts from the 3 tiny samples (\SIrange{0.1}{0.3}{\percent}).
Figure~\ref{fig:size-prediction} shows the error of \papertitle{} in predicting the size of cached datasets. 
On average, the error is \SI{7.4}{\percent}.
We observe the best case in \textsc{svm} (\SI{0.0008}{\percent}) and the worst case in \textsc{gbt} (\SI{36.7}{\percent}). 
With the exception of \textsc{gbt}, we see the high accuracy of \papertitle{} -- higher than \SI{85}{\percent} in \textsc{als} and \SI{95}{\percent} in all remaining applications. 

\smallskip

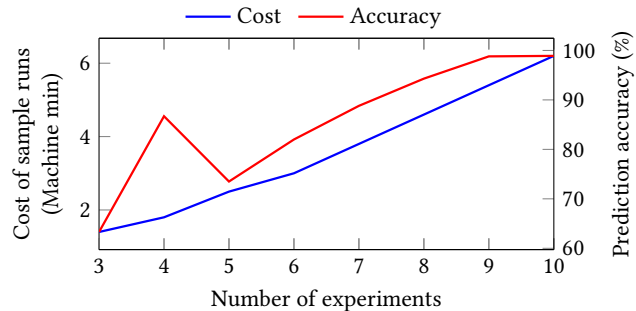
\begin{figure}[t!]
\centering
\begin{tikzpicture}

 \begin{axis}[
    legend style={at={(0.3,1.2)},
    draw=none,
    /tikz/every even column/.append style={column sep=0.5cm},
    anchor=north,
    legend columns=-1},
    height=125pt,
    width=0.9\columnwidth,
    axis y line*=left,
    xmin=3,
    xmax=10,
    xlabel=Number of experiments,
    ylabel style={align=center},
    ylabel=Cost of sample runs\\(Machine min),
    xtick=data,
    ]
    \addplot[line width=0.3mm, color=blue] table[x=Experiments, y=SamplingCost, col sep=comma] {data/gbt-sampling-cost-vs-prediction-accuracy-ratio.dat}; 
    \label{pgfplots:gbt-sampling-cost}
    \legend{Cost}
 \end{axis}
     
 \begin{axis}[
   legend style={at={(0.6,1.2)},
   draw=none,
   /tikz/every even column/.append style={column sep=0.5cm},
   anchor=north,
   legend columns=-1},
   height=125pt,
   width=0.9\columnwidth,
   axis y line*=right,
   xmin=3,
   xmax=10,
   ylabel style={align=center},
   ylabel=Prediction accuracy (\%),
   xtick=data,
   xticklabels=\empty,
   ]
   \addplot[line width=0.3mm, color=red] table[x=Experiments, y=PredictionAccuracy, col sep=comma] {data/gbt-sampling-cost-vs-prediction-accuracy-ratio.dat}; 
   \label{pgfplots:gbt-prediction-accuracy}
   \legend{Accuracy}
 \end{axis}

\end{tikzpicture}
\vspace{-25pt}
\caption{Cost of sample runs and prediction accuracy (\textsc{gbt}).}
\vspace{-20pt}
\label{fig:gbt-sampling-cost-vs-prediction-accuracy-ratio}
\end{figure}

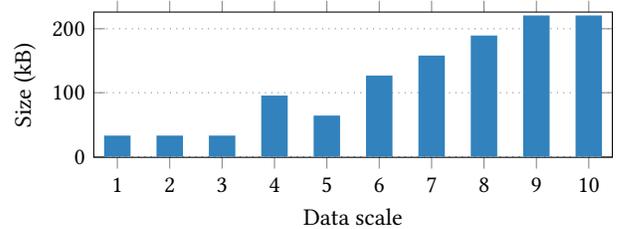
\begin{figure}[t!]
\centering
\begin{tikzpicture}

 \begin{axis}[
 	     ybar,
             bar width=10pt,
             xtick=data,
             ymin=0,
             ymax = 225,
             ylabel=Size (\SI{}{\kilo\byte}),
             xlabel=Data scale,
             width=1.0\columnwidth,
             enlarge x limits = 0.05,
             enlarge y limits = 0.005,
             height=100pt,
             ymajorgrids,
             table/col sep=comma,
             table/x=Datascale,
             cycle list/Blues-3,
             every axis plot/.append style={fill},
             legend style={at={(0.5,1.5)},
             draw=none,
             /tikz/every even column/.append style={column sep=0.5cm},
             anchor=north,
             legend columns=-1},
             ]
	    \addplot[draw=none, fill=Blues-I] table [x=Datascale, y=DataSizeKB] {data/gbt-data-size.dat};
	    \label{pgfplots:gbt-data-size}
 \end{axis}

\end{tikzpicture}
\vspace{-10pt}
\caption{Size of cached datasets during sample runs (\textsc{gbt}).}
\vspace{-15pt}
\label{fig:gbt-data-size}
\end{figure}

\noindent
\textbf{\textsc{gbt.}}
We discuss the relatively low size prediction accuracy for \textsc{gbt} (\SI{63.3}{\percent}). 
To get better insights, we additionally run 7 sample experiments with new data scales (\SIrange{0.4}{1.0}{\percent} of the original data) and plot the change in cost of sample runs \& prediction accuracy with an increase in the number of sample runs. 
Figure~\ref{fig:gbt-sampling-cost-vs-prediction-accuracy-ratio} shows that while the cost of sample runs increases, the prediction accuracy improves when we add more sample runs. 
For 10 sample runs, prediction accuracy is \SI{98.9}{\percent}. 
This indicates that the poor prediction accuracy of the size of cached datasets is due to insufficient sample runs, not due to the selected model (Equation~\ref{eq:size-prediction}). 
In spite of data size prediction error, \papertitle{} selects the optimal cluster size (i.e., a single machine) because both the predicted size (\SI{13.8}{\mega\byte}) and the actual size (\SI{21.7}{\mega\byte}) fit the memory of a single machine. 
Another explanation for the poor accuracy in \textsc{gbt} is the small size of the original data (\SI{30.6}{\mega\byte}).
Therefore, during the 3 sample runs, the training data is only a few kilobytes.
Figure~\ref{fig:gbt-data-size} shows the size of cached datasets in sample runs with small data scales. 
These are used to train the size prediction models. 
Using \emph{cross validation} with 3 sample runs to evaluate 
the models, we get a model error of \SI{53.9}{\percent}. 
With 10 sample runs however, we get an error of \SI{28.5}{\percent}.
Since this error can be measured by the sample runs manager, we leave it as a future work for \papertitle{} to apply \emph{adaptive sampling} by carrying out additional sample runs to limit the error to a predefined threshold.
\subsection{Overhead of sample runs}
\begin{figure}[t!]
\centering
\begin{tikzpicture}
\begin{axis}[ybar,
             bar width=8pt,
             xtick=data,
             ymin=0,
             ymax=130,
             ymode = log,
             ytick={3, 7, 15, 30, 60, 120},
             yticklabels={3, 7, 15, 30, 60, 120},
             ylabel=Cost (\%),
             width=1.0\columnwidth,
             enlarge x limits = 0.05,
             enlarge y limits = 0.005,
             height=100pt,
             ymajorgrids,
             table/col sep=comma,
             table/x=Application,
             xtick={0,...,8},
             cycle list/Blues-3,
             every axis plot/.append style={fill},
             xticklabels={{\textsc{als}},{\textsc{bayes}},{\textsc{gbt}},{\textsc{km}},{\textsc{lr}},{\textsc{pca}},{\textsc{rfc}},{\textsc{svm}}},
             legend style={at={(0.5,1.31)},
             draw=none,
             /tikz/every even column/.append style={column sep=0.5cm},
             anchor=north,
             legend columns=-1},
             ]
\addplot[draw=none, fill=Blues-F] table [x expr=\coordindex, y=TrainingVersusOptimalRatio] {data/sampling-vs-optimal-cost-ratio.dat};
\label{pgfplots:Blink-vs-optimal-cost-ratio}
\addplot[draw=none, fill=Blues-I] table [x expr=\coordindex, y=ErnestVersusOptimalRatio] {data/Ernest-vs-optimal-cost-ratio.dat};
\label{pgfplots:Ernest-vs-optimal-cost-ratio}
\legend{Blink, Ernest}
\end{axis}
\end{tikzpicture}
\vspace{-10pt}
\caption{Cost of sample runs of \papertitle{} and Ernest compared to cost of optimal actual runs.}
\vspace{-20pt}
\label{fig:sampling-to-optimal-cost-ratio}
\end{figure}
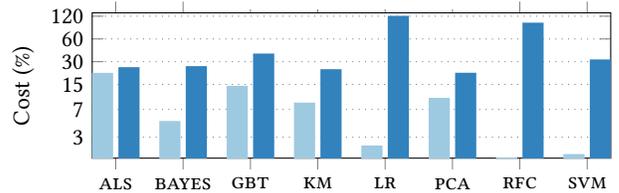

To evaluate the overhead of sample runs, we compare their cost with the cost of the corresponding actual run on optimal cluster configuration. 
Figure~\ref{fig:sampling-to-optimal-cost-ratio} shows that on average, sample runs of an application cost \SI{8.1}{\percent} compared with the cost of its actual run on optimal cluster size. 
At worst, the overhead is \SI{21.3}{\percent} (\textsc{als}) while at best, it is \SI{1.6}{\percent} (\textsc{rfc}).
Taking each sampling approach separately, we see that the average cost of sample runs of \bnumber{} is \SI{2.7}{\percent}, with a worst case of \SI{5.1}{\percent} (\textsc{bayes}) and a best case of \SI{1.6}{\percent} (\textsc{rfc}). 
For \bsize{}, the average cost of sample runs is \SI{13.3}{\percent}, with a worst case of \SI{21.3}{\percent} (\textsc{als}) and a best case of \SI{8.6}{\percent} (\textsc{km}). 
Altogether, \bsize{} costs about $ 4.9\times $ more than \bnumber{}. 
Nonetheless, the cost of \bsize{} is still tolerable because we are comparing its cost with costs of optimal actual runs. 
It is worth mentioning that all sample runs are carried out without changing any application parameter (e.g., number of iterations).
Taking \textsc{km} as a short-running application (\SI{3.5}{\minute} on the optimal cluster size; cf.~Table~\ref{tbl:application-details}), 
sample runs cost \SI{8.6}{\percent} of the cost of the actual run on the optimal cluster size. Hence, \papertitle{} is also effective for short-running applications.

Even though Ernest (cf.~Section~\ref{sec:related-work}) predicts application runtime rather than cluster size, we compare the cost of its sample runs with the cost of those carried out by the sample runs manager (cf.~Section~\ref{subsec:light-weight-sampling}) to see how much cost savings can be achieved with the efficient sample runs applied in \papertitle{}.
We carry out 7 sample runs, as recommended by Ernest's \emph{optimal experiment design}~\cite{Pukelsheim:1993:OptimalDoE}, on \SIrange{1}{12}{} machines with sample datasets (\SIrange{1}{10}{\percent} of the original data).
The sample runs of Ernest cost 16.4$\times$ more than those of \papertitle{} (as depicted in Figure~\ref{fig:sampling-to-optimal-cost-ratio}) and 3.8$\times$ more than Cherrypick (as stated in \cite{Alipourfard:2017:NSDI:CherryPick}).
Thus, the cost of sample runs of \papertitle{} is cheaper than those of both Ernest and Cherrypick.

\subsection{Scalability}
\label{subsec:data-sample-reduction}

After demonstrating the efficiency of \papertitle{} using experimental evaluations based on 3 sample runs (\SI{0.1}{\percent} - \SI{0.3}{\percent} of the original data), we evaluate the accuracy of \papertitle{} in selecting optimal cluster size with larger input data sizes.
In other words, we want to answer questions like: 
\emph{Will \papertitle{} still be effective when we use smaller sample data scales (less than \SI{0.1}{\percent})?} 
\emph{Can the same extracted size prediction models be reused for larger input data (more than \SI{100}{\percent})?} 
To answer these questions, 
we increase the size of input data (i.e., \SI{150}{\percent} - \SI{18e4}{\percent} instead of \SI{100}{\percent}) and reuse the previously extracted size prediction models to predict the optimal cluster sizes. 

As an exception, to predict precisely the size of cached datasets in \textsc{als} and \textsc{gbt}, we conducted 5 and 10 sample runs respectively (cf.~Section~\ref{subsec:prediction-accuracy-of-cached-intermediate-results}).
We can see from Table~\ref{tbl:application-details} (with data scale +\SI{150}{\percent}) that \papertitle{} still selects the optimal cluster size for all applications except \textsc{km}. 
Comparing the cost of sample runs with the cost of actual runs on optimal cluster size, we see that the cost is \SI{1.08}{\percent} on average, \SI{2.5}{\percent} in the worst case (\textsc{Bayes}) and \SI{0.2}{\percent} in the best case (\textsc{pca}). 
Although \papertitle{} made 10 sample runs for \textsc{gbt}, their cost is \SI{2.35}{\percent} compared with the actual run on optimal cluster size. 
It is worth highlighting cases like \textsc{pca} and \textsc{gbt} where sampling data size to actual data size ratio are (\SI{2e-3}{\percent}-\SI{6e-3}{\percent}) and (\SI{6e-5}{\percent}-\SI{6e-4}{\percent}) respectively.

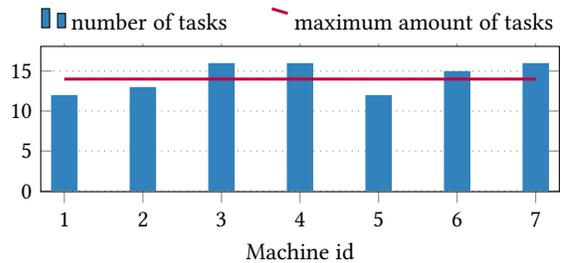
\begin{figure}[t!]
\centering
\begin{tikzpicture}

 \begin{axis}[
 	     ybar,
             bar width=10pt,
             xtick=data,
             ymin=0,
             ymax = 18,
             xlabel=Machine id,
             width=1.0\columnwidth,
             enlarge x limits = 0.05,
             enlarge y limits = 0.005,
             height=100pt,
             ymajorgrids,
             table/col sep=comma,
             table/x=machine,
             cycle list/Blues-3,
             every axis plot/.append style={fill},
             legend style={at={(0.5,1.31)},
             draw=none,
             /tikz/every even column/.append style={column sep=0.5cm},
             anchor=north,
             legend columns=-1},
             ]
	    \addplot[draw=none, fill=Blues-I] table [x=machine, y=tasks] {data/km-task-distribution.dat};
	    \label{pgfplots:number-of-tasks}
	    
	    \addplot[line width=0.4mm, purple,sharp plot] table [y=line] {data/km-task-distribution.dat};
        \label{pgfplots:maximum-tasks}
        
        \legend{number of tasks, maximum amount of tasks}
 \end{axis}
 
\end{tikzpicture}
\vspace{-10pt}
\caption{Tasks distribution among machines (\textsc{km}).}
\vspace{-10pt}
\label{fig:TasksDistribution-KMeans}
\end{figure}

\noindent \textbf{K-means.} 
\papertitle{} predicts the size of cached datasets in \textsc{km} with an accuracy of \SI{99.7}{\percent}.
However, despite the high prediction accuracy, \papertitle{} selects 7 machines for \textsc{km}, which costs \SI{104.3}{} machine minutes, instead of 8 machines, which is the optimal cluster size with a cost of \SI{36.4} machine minutes, as shown in Table~\ref{tbl:application-details} (with data scale +\SI{150}{\percent}). 
The reason behind the selection of suboptimal cluster size is the unequal distribution of tasks among machines. 
The application parallelism is 100 and the number of recommended machines is 7. Thus, the capacity of each machine is 14 tasks. 
However, some machines run more than 14 tasks and therefore some cached data partitions get evicted. 
Figure~\ref{fig:TasksDistribution-KMeans} illustrates the distribution of tasks in \textsc{km} on 7 machines. 
We observe that 7 cached data partitions are evicted from memory, which is the same number of tasks that are over-assigned to machines (i.e., 1, 2, 2, and 2 tasks in machines $6$, $3$, $4$ and $7$, respectively).

\subsection{Cluster bounds}
\begin{table}[tb]
\centering
\small{
\npdecimalsign{.}
\nprounddigits{2}
\vspace{5pt}
 \begin{adjustbox}{max width=\columnwidth}
 \begin{tabular}{|c | c | c | c | c | c | c | c |} 
\hline 
 Scale~\textbackslash~App & \textsc{als} & \textsc{Bayes} & \textsc{gbt} & \textsc{lr} & \textsc{pca} & \textsc{rfc} & \textsc{svm} \\  
\hline 
 -5\% & \cmark & \cmark & \cmark & \cmark & \cmark & \cmark & \cmark \\ 
 \hline
 -4\% & \cmark & \cmark & \xmark & \cmark & \xmark & \cmark & \cmark \\ 
 \hline
 -3\% & \cmark & \cmark & \xmark & \cmark & \xmark & \xmark & \cmark \\ 
 \hline
 -2\% & \cmark & \cmark & \xmark & \cmark & \xmark & \xmark & \xmark \\ 
 \hline
 -1\% & \xmark & \cmark & \xmark & \cmark & \xmark & \xmark & \xmark \\
 \hline
 Predicted & \xmark & \cmark & \xmark & \cmark & \xmark & \xmark & \xmark \\ 
 \hline
 +1\% & \xmark & \cmark & \xmark & \cmark & \xmark & \xmark & \xmark \\
 \hline
 +2\% & \xmark & \xmark & \xmark & \cmark & \xmark & \xmark & \xmark \\
 \hline
 +3\% & \xmark & \xmark & \xmark & \cmark & \xmark & \xmark & \xmark \\
 \hline
 +4\% & \xmark & \xmark & \xmark & \cmark & \xmark & \xmark & \xmark \\
 \hline
 +5\% & \xmark & \xmark & \xmark & \xmark & \xmark & \xmark & \xmark \\
\hline 
 \end{tabular}
 \end{adjustbox}
\vspace{-5pt} 
\caption{Cluster bounds. \cmark refers to eviction-free runs.}
\label{tbl:cluster-bounds}
\vspace{-10pt}
}
\end{table}
In addition to selecting the optimal cluster size, we also evaluate \papertitle{} with respect to predicting the maximum input data scale that guarantees eviction-free runs on a resource-constrained cluster. 
We fix the cluster size at 12 machines and predict for each application the maximum input data scale that the cluster runs efficiently.
We identify whether \papertitle{'s} predicted data scale for each application is larger or smaller than the maximum data scale for an eviction-free run of the application and measure the difference. 
Table~\ref{tbl:cluster-bounds} shows the prediction accuracy of \papertitle{} for all the 7 applications, after excluding \textsc{km} (cf.~Section~\ref{subsec:data-sample-reduction}). 
\papertitle{} predicts, with a tolerance of \SI{\pm 5}{\percent}, the input data scale for which the applications would execute without eviction. 
It can be seen that \textsc{lr} has the maximum upper bound for which eviction is guaranteed not to occur -- \SI{4}{\percent} larger than the predicted scale by \papertitle{}. 
While \textsc{gbt} and \textsc{pca} have the minimum upper bound for which eviction is guaranteed not to occur -- \SI{5}{\percent} smaller than the predicted scale by \papertitle{}.

\section{Conclusion}
\label{sec:conclusion}

\papertitle{} is an autonomous sampling-based framework that selects an optimal cluster size with the highest cost efficiency for running non-recurring iterative big data applications. 
Overall, the evaluation of \papertitle{} shows very good results in terms of selecting an optimal cluster size with high prediction accuracy using lightweight sample runs.
The evaluation also proves the re-usability of sample runs in new cluster environments and very large data scales.

\textbf{Acknowledgement}
This research was partially funded by the Thuringian Ministry for Economy, Science and Digital Society under the project thurAI and by the Carl-Zeiss-Stiftung under the project “Memristive Materials for Neuromorphic Electronics (MemWerk)".
\balance

\clearpage

\bibliographystyle{ACM-Reference-Format}
\bibliography{bibliography/main}

\end{document}